\documentclass[aps,,showpacs,twocolumn,superscriptaddress,floatfix]{revtex4}
\usepackage{graphicx}
\usepackage{dcolumn} 

\begin{document} 

\title{Three-body calculation of the $\Delta\Delta$ dibaryon candidate 
${\cal D}_{03}(2370)$} 

\author{A.~Gal\footnote{Corresponding author: avragal@vms.huji.ac.il}}
\affiliation{Racah Institute of Physics, The Hebrew University, 
Jerusalem 91904, Israel}

\author{H.~Garcilazo\footnote{humberto@esfm.ipn.mx}} 
\affiliation{Escuela Superior de F\' \i sica y Matem\'aticas, 
Instituto Polit\'ecnico Nacional, Edificio 9, 
07738 M\'exico D.F., Mexico}

\date{\today}

\begin{abstract}
\rule{0ex}{3ex} 

The ${\cal D}_{03}$ dibaryon is generated dynamically as a resonance pole 
in a $\pi N \Delta'$ three-body model, where $\Delta'$ is a stable $\Delta$ 
baryon. Using separable interactions dominated by the $\Delta(1232)$ isobar 
for $\pi N$ and by the ${\cal D}_{12}(2150)$ isobar for $N\Delta'$, with 
${\cal D}_{12}(2150)$ the $N\Delta$ dibaryon deduced in and constrained by 
$^1D_2$ $pp$ scattering, the model reduces to an effective two-body problem 
for $\Delta\Delta'$ which is solved. The mass and width of ${\cal D}_{03}$ 
are found close to those of the $I(J^P)=0(3^+)$ resonance peak observed by 
WASA@COSY in pion-production $pn$ collisions at 2.37 GeV. 

\end{abstract}

\maketitle

\noindent 
{\bf Introduction.}~~
Arguments in favor of $N\Delta$ and $\Delta\Delta$ dibaryon resonances and 
estimates of their mass values relative to the respective thresholds at 2.17 
and 2.46 GeV  date back to 1964 \cite{dyson64}, as soon as SU(6) symmetry 
proved useful in classifying baryons and mesons below 2 GeV. Since nucleons 
and $\Delta$'s belong to the $\bf{56}$ representation, product states of 
$\bf{56 \times 56}$ offer numerous nonstrange dibaryon candidates. Focusing 
on $L=0$ $s$-wave dibaryons ${\cal D}_{IS}$, with isospin $I$ and spin $S$, 
and on the $\bf{\overline{10}}$ and $\bf{27}$ SU(3) multiplets that contain 
the deuteron ${\cal D}_{01}$ and $NN$ virtual state ${\cal D}_{10}$, 
respectively, these symmetry-based arguments leave only two additional 
nonstrange dibaryon candidates: ${\cal D}_{12}$ and ${\cal D}_{03}$ with 
predicted masses 2.16 and 2.35 GeV, respectively \cite{dyson64}.

Of these two $s$-wave dibaryon candidates, the ${\cal D}_{12}$ shows up 
experimentally as an $NN({^1D_2})$ $\leftrightarrow$ $\pi d({^3P_2})$ 
coupled-channel resonance corresponding to a quasibound $N\Delta$ with 
mass 2.15 GeV, near the $N\Delta$ threshold, and width about 0.115 GeV 
\cite{hoshizaki78,arndt81}. Early versions of quark models placed 
${\cal D}_{12}$ almost 200 MeV too high \cite{mulders80,mulders83}, 
but subsequent chiral quark cluster model $N\Delta$ calculations place it 
at 2.17 GeV, right on threshold \cite{mota99}. Elsewhere we show in detail 
that ${\cal D}_{12}$ also appears within a dynamical $\pi NN$ three-body 
model as a robust $N\Delta$ dibaryon resonance, 
with $M-i\;(\Gamma/2)\approx 2.15-i\;0.06$ GeV \cite{gg14}, 
compatible with its observed mass and width. 

Experimental evidence for ${\cal D}_{03}$ developed in the 1970s by 
observing a resonance-like behavior of the proton polarization in 
$\gamma d \to pn$ at $\sqrt{s}\approx 2.38$ GeV and correlating it 
with a strong $\Delta\Delta$ attraction using quark-model coupling 
constants in one-boson-exchange (OBE) model calculations \cite{kamae77}. 
Subsequent OBE calculations \cite{SS83} and quark-model based calculations 
\cite{mulders80,mulders83,oka80,maltman85,gold89,garcilazo97} of the (real) 
$\Delta\Delta$ interaction yielded binding energies ranging from a few to 
hundreds of MeV. The most recent evidence for ${\cal D}_{03}$, with mass 
2.37 GeV, comes from exclusive and kinematically-complete high-statistics 
measurements of $np\to d\pi^0\pi^0, d\pi^+\pi^-$ two-pion production reactions 
by WASA@COSY \cite{bash09}. A most intriguing feature of this resonance, 
particularly when interpreted as a $\Delta\Delta$ dibaryon bound by 90 MeV, 
is the width of $\Gamma\approx 70$~MeV which is remarkably smaller than given 
by a naive estimate $\Gamma_{\Delta}\lesssim\Gamma\lesssim 2\Gamma_{\Delta}$, 
where $\Gamma_{\Delta}\approx 120$ MeV. 

In this Letter we study the ${\cal D}_{03}$ dibaryon using nucleons and pions 
as the relevant hadronic degrees of freedom, rather than constituent quarks 
used in most past works to generate $N\Delta$ and $\Delta\Delta$ interaction 
potentials for solving a two-body (mostly Schroedinger) wave equation. Given 
the prominent role of the $\Delta$(1232) resonance in $\pi N$ dynamics, we 
extend our hadronic building blocks to include $\Delta$'s as self-consistently 
as possible. Earlier attempts by Ueda to consider dibaryon resonances within 
$\pi NN$ and $\pi\pi NN$ dynamics were limited to Heitler-London estimates 
\cite{ueda78}, followed by $\pi NN$ Faddeev calculations with unrealistic 
nonrelativistic kinematics \cite{ueda82}. In contrast, we solve three-body 
Faddeev equations with relativistic kinematics, $\pi NN$ for ${\cal D}_{12}$ 
and $\pi N\Delta$--like for ${\cal D}_{03}$, the latter substituting for 
$\pi\pi NN$ four-body Faddeev-Yakubovsky equations. Each of these derived 
dibaryon resonances is the lowest, and perhaps the only $s$-wave dibaryon 
within its own class. Here we focus on ${\cal D}_{03}$ in the first 
quantitative phenomenological few-body calculation to confront realistically 
the recently observed WASA@COSY 2.37 GeV resonance peak \cite{bash09}. 

\noindent 
{\bf Three-body model of ${\cal D}_{03}$.}~~~In Table~\ref{tab:thresholds} 
we list two-body thresholds relevant for cluster decompositions of the 
$I(J^P)=0(3^+)$ $\pi\pi NN$ system. At 1~fm separation distance, the $\ell=2$ 
centrifugal energy upward shifts of at least 200 MeV make the channels 
$(NN)_d$--$(\pi\pi)_{\sigma(500)}$ and $N$--$(\pi\pi N)_{N^{\ast}(1440)}$ 
incompetitive against the $\Delta$--$\Delta$ channel with $\ell=0$ 
threshold at 2460 MeV. For $\pi$--$(\pi NN)_{{\cal D}_{12}(2150)}$, the 
$\Delta$-dominated $\pi N$ interaction should reduce the $\ell=1$ upward 
shift from 300 MeV (no interaction) to about 150 MeV, so that the effective 
threshold here becomes somewhat lower than the $\ell=0$ $\Delta\Delta$ 
threshold. This singles out $\Delta(1232)$ and ${\cal D}_{12}(2150)$ as the 
most likely fermionic degrees of freedom of which pions may benefit in forming 
the $I(J^P)=0(3^+)$ $\pi\pi NN$ system, with ${\cal D}_{03}$ likely to emerge 
as a quasibound state in $(\Delta\Delta)_{\rm upper}$--$[\pi{\cal D}_{12}
(2150)]_{\rm lower}$ coupled-channel calculations. This is consistent 
with the observation that ${\cal D}_{03} \to {\cal D}_{12}+\pi$ provides 
a dominant doorway decay mode into two-pion final states \cite{kukulin13}. 
We also note that by assigning $s$-wave $N\Delta$ structure to 
${\cal D}_{12}$, both angular-momentum and isospin recoupling coefficients 
for transforming $\pi{\cal D}_{12}(2150)$ with $p$-wave pion in a $I(J^P)=
0(3^+)$ state into $s$-wave $\Delta\Delta$ are equal to 1, thus maximizing 
the coupling between these channels. 

\begin{table}[t] 
\caption{Two-body threshold energies $E_{\rm th}$ (in MeV) and lowest partial 
waves $\ell$ for the $I(J^P)=0(3^+)$ $\pi\pi NN$ system.} 
\begin{ruledtabular} 
\begin{tabular}{ccccc} 
& $\Delta$--$\Delta$ & $d$--$\sigma(500)$ & $N$--$N^{\ast}(1440)$ & 
$\pi$--${\cal D}_{12}(2150)$  \\ 
\hline  
$E_{\rm th}$ & 2460 & 2380 & 2380 & 2290  \\ 
$\ell$ & 0 & 2 & 2 & 1  \\ 
\end{tabular} 
\end{ruledtabular}
\label{tab:thresholds} 
\end{table} 

\begin{figure}[hbt] 
\centerline{\includegraphics[scale=0.45]{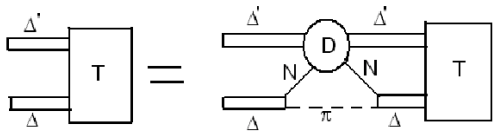}} 
\caption{Graphical representation of the $\Delta\Delta'$ dibaryon 
$T$-matrix pole equation. $D$ denotes the ${\cal D}_{12}$ isobar.} 
\label{fig:DelDel'} 
\end{figure}

These arguments led us to reduce the $I(J^P)=0(3^+)$ $\pi\pi NN$ 
system to a system of three hadrons $\pi$, $N$, $\Delta'$ interacting 
via pairwise separable potentials. This approximates one of the $\pi N$ 
resonating pairs in the four-body system by a {\it stable} $\Delta$ of 
mass 1232 MeV and zero width, here denoted $\Delta'$. In this model the 
$\pi N$ interaction is limited to the $I(J^P)=\frac{3}{2}({\frac{3}{2}}^+)$ 
$p$-wave channel (denoted $P_{33}$) dominated by the $\Delta$ resonance. 
The $N\Delta'$ interaction is limited to the $I(J^P)=1(2^+)$ $s$-wave 
channel dominated by the ${\cal D}_{12}$ dibaryon and, finally, the 
$\pi\Delta'$ interaction is neglected because the mass of the lightest 
$N^{\ast}(\frac{5}{2}^+)$ isobar candidate $N^{\ast}(1680)$ is too high for 
our purpose. This $\pi N\Delta'$ three-body model leads to a $\Delta\Delta'$ 
eigenvalue problem for the $T$ matrix diagram shown in Fig.~\ref{fig:DelDel'}, 
where starting with $\Delta\Delta'$, the $\Delta$ resonance isobar decays 
into a $\pi N$ pair followed by $N\Delta'\to N\Delta'$ scattering via the 
${\cal D}_{12}$ isobar (marked $D$ in the figure) with a spectator pion, 
and finally by $\pi N\to\Delta$ fusion back into the $\Delta\Delta'$ channel. 

We note that, whereas the $\Delta$-isobar decay to the $\pi N$ channel is 
fully accounted for, the ${\cal D}_{12}$ isobar is allowed to decay only to 
$N\Delta'$. Additional width contribution will arise upon allowing the stable 
$\Delta'$ in the $\pi N\Delta'$ model to acquire normal $\Delta$$\to$$N\pi$ 
decay width. However, this added $\Delta'$ width is partly suppressed by 
quantum-statistics correlations between the decay $N\pi$ pair and the 
pre-existing $N\pi$ pair. Thus, for $s$-wave nucleons and $p$-wave pions, 
$J^P$=$3^+$ implies space-spin symmetry for nucleons as well as for pions. 
With total $I$=0, Fermi-Dirac (Bose-Einstein) statistics for nucleons 
(pions) allows for isospins $I_{NN}$=$I_{\pi\pi}$=0, forbidding 
$I_{NN}$=$I_{\pi\pi}$=1, 
with weights 2/3 and 1/3, respectively, obtained by recoupling the two 
isospins $I_{N\pi}$=3/2 in the $I$=0 $\Delta\Delta$ state. 

We now specify the pairwise interactions in the $\pi N\Delta'$ three-body 
model. The interaction between particles 1,2 is denoted $V_3$, etc. Since 
the $\pi N$ interaction $V_3$ is dominated by the $\Delta(1232)$ isobar 
resonance, it is limited here to a $P_{33}$ rank-one separable-potential 
of the form 
\begin{equation} 
V_3(p_3,p'_3)=\lambda_3 g_3(p_3)g_3(p_3^\prime),
\label{eq:V3}
\end{equation}
so that the corresponding $t$ matrix is given by 
\begin{equation}
t_3(\omega_3;p_3,p_3^\prime)= g_3(p_3)\tau_3(\omega_3)g_3(p_3^\prime),
\label{eq:t3}
\end{equation}
where $\tau_3(\omega_3)$ is the propagator of the $\Delta(1232)$ isobar 
in the two-body cm system, with $\omega_3$ the two-body $\pi N$ cm energy. 
In the three-body cm system, with $W$ the total three-body cm energy and 
$q_3$ the momentum of the spectator $\Delta'$ with respect to the two-body 
$\pi N$ isobar, it assumes the form 
\begin{equation} 
{\cal T}_3^{-1}(W;q_3)=\lambda_3^{-1}-\int_0^\infty
\frac{[g_3(p_3)]^2\; p_3^2\; dp_3}{W-E_{\Delta'}(q_3)-
{\cal E}_3(p_3,q_3)+i\epsilon}, 
\label{eq:calT3} 
\end{equation} 
where ${\cal E}_3(p_3,q_3)=[(E_{\pi}(p_3)+E_N(p_3))^2+q_3^2]^{\frac{1}{2}}$, 
with $E_h(p)$=$(m_h^2+p^2)^{\frac{1}{2}}$ for hadron $h$. 
For $q_3=0$, when the three-body cm system degenerates to the two-body 
cm system, ${\cal T}_3$ and $\tau_3$ are related by a simple mass shift, 
${\cal T}_3(W;q_3=0)=\tau_3(W-m_{\Delta'})$, as expected. 

\begin{table}[t] 
\caption{$\pi N$ $P_{33}$ form factor $g_3(p)$ parameters [Eqs.~(\ref{eq:V3}), 
(\ref{eq:g31}), (\ref{eq:g32})], and the zero $r_0$ of the Fourier transform 
${\tilde g}_3(r)$ \cite{gg11}.} 
\begin{ruledtabular}
\begin{tabular}{cccccc} 
$k$ & $\lambda_3\;({\rm fm}^4)$ & $\beta\;({\rm fm}^{-1})$ & 
$\gamma\;({\rm fm}^{-1})$ & $C\;({\rm fm}^2)$ & $r_0 \;({\rm fm})$  \\ 
\hline 
1 & $-$0.07587 & 1.04 & 2.367 & 0.23 & 1.36  \\ 
2 & $-$0.04177 & 1.46 & 4.102 & 0.11 & 0.91  \\
\end{tabular}
\end{ruledtabular}
\label{tab:piN} 
\end{table} 

For the $\pi N$ $p$-wave form factor $g_3$ we considered two forms, labeled 
here $k$=1,2: 
\begin{equation} 
g_3(p_3)=p_3\exp(-p_3^2/\beta^2)+Cp_3^3\exp(-p_3^2/\gamma^2),\;\;\;(k=1) 
\label{eq:g31} 
\end{equation}
falling off exponentially \cite{gg11}, and 
\begin{equation} 
g_3(p_3)=\frac{p_3}{(1+p_3^2/\beta^2)^2}+C\frac{p_3^3}
{(1+p_3^2/\gamma^2)^3},\;\;\;(k=2) 
\label{eq:g32} 
\end{equation} 
falling-off as inverse power of $p_3$, here as $p_3^{-3}$. Both form 
factors, using the parameters listed in Table~\ref{tab:piN}, reproduce 
perfectly the $\pi N$ $P_{33}$ phase shifts from Ref.~\cite{arndt06}. 
The table also lists the distance $r_0$ at which the Fourier transform 
${\tilde g}_3(r)$ flips sign, which roughly represents the spatial extension 
of the $P_{33}$ $p$-wave form factor \cite{gg11}. Together with a rank-two 
separable $NN$ potential reproducing the $^3S_1$ phase shift, these form 
factors lead in a relativistic $\pi NN$ Faddeev calculation to ${\cal D}_{12}$ 
dibaryon pole at $E=M-i\;(\Gamma/2)=2151(2)-i\;60(3)$ MeV 
\cite{gg14}, in good agreement with accepted values \cite{hoshizaki78}. 

\begin{figure}[thb] 
\centerline{\includegraphics[scale=0.35]{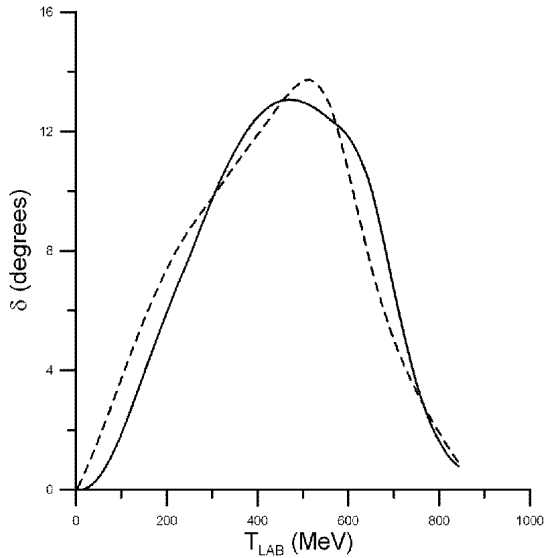}} 
\caption{The $^1D_2$ $NN$ phase shift $\delta$. Dashed: Arndt {\it et al.} 
\cite{arndt07}. Solid: Eq.~(\ref{eq:g1}) best fit for $A_1=A_2=A_3=1$.} 
\label{fig:del} 
\end{figure} 

The $N\Delta'$ interaction $V_1$, dominated by the ${\cal D}_{12}(2150)$ 
isobar resonance, is limited here to the $I(J^P)=1(2^+)$ channel. 
${\cal D}_{12}$ shows up as an inelastic resonance in the $^1D_2$ $NN$ 
partial-wave above the $\pi NN$ threshold \cite{hoshizaki78,arndt07}. 
To generate the necessary inelastic $\pi NN$ cut we introduced a third 
$s$-wave subchannel $NN'$ where $N'$ is a fictitious nonstrange stable 
baryon with $I(J^P)=\frac{1}{2}(\frac{3}{2}^+)$ and $m_{N'}=m_N+m_\pi$. We 
have then fitted the $NN$ $^1D_2$ partial-wave amplitude of Arndt {\it et al.} 
\cite{arndt07} using a coupled-channel separable potential  
\begin{equation}
V_1^{mn}(p_1,p_1')= \lambda_1 g_1^m(p_1) g_1^n(p_1') \,\,\,\,\,\,\,
(m,n=1-3),
\label{eq:V1}
\end{equation}
where the three subchannels labeled $m,n$ are 1=$NN$ ($d$-wave), 2=$NN'$ 
($s$-wave), and 3=$N\Delta'$ ($s$-wave). The $t$-matrix of the system is 
obtained by solving a Lippmann-Schwinger equation with relativistic kinematics 
which in the case of the separable potential (\ref{eq:V1}) has the solution 
\begin{equation} 
t_1^{mn}(\omega_1;p_1,p_1')= g_1^m(p_1)\tau_1(\omega_1)g_1^n(p_1^\prime),
\label{eq:t1}
\end{equation}
where $\tau_1(\omega_1)$ is the propagator of the ${\cal D}_{12}$ isobar 
in the two-body cm system. In the three-body cm system it assumes the form 
\begin{equation}
{\cal T}_1^{-1}(W;q_1)=\lambda_1^{-1}-\sum_{r=1}^3 \int_0^\infty 
\frac{[g_1^r(p_1)]^2\; p_1^2\; dp_1}
{W-E_{\pi}(q_1)-{\cal E}_1^r(p_1,q_1)+i\epsilon},
\label{eq:calT1}
\end{equation} 
where ${\cal E}_1^r(p_1,q_1)=[(E_N(p_1)+E_r(p_1))^2+q_1^2]^{\frac{1}{2}}$, 
with masses $m_r=(m_N,m_{N'},m_{\Delta'})$ for $r=(1,2,3)$. 

\begin{figure}[thb] 
\centerline{\includegraphics[scale=0.35]{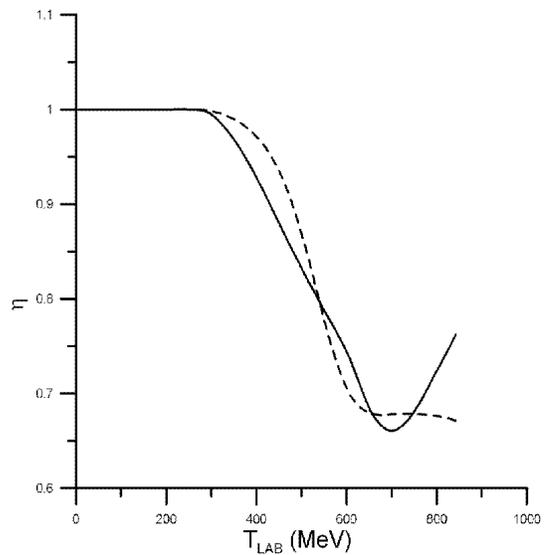}} 
\caption{The $^1D_2$ $NN$ inelasticity $\eta$. Dashed: Arndt {\it et al.} 
\cite{arndt07}. Solid: Eq.~(\ref{eq:g1}) best fit for $A_1=A_2=A_3=1$.}
\label{fig:eta}
\end{figure} 

The form factors of the separable potential (\ref{eq:V1}) were taken in the 
following form: 
\begin{equation}
g_1^n(p_1)=\frac{p_1^\ell}{[1+p_1^2/(\alpha_n)^2]^{1+\frac{\ell}{2}}}
\left( 1+A_n\frac{p_1^2}{1+p_1^2/(\alpha_n)^2} \right),
\label{eq:g1}
\end{equation}
where $\ell=2$ for $n=1$, and $\ell=0$ for $n=2,3$. The range parameters 
$\alpha_n$ were limited to values $\alpha_n \lesssim 3$~fm$^{-1}$ to 
ensure that the physics of these coupled channels does not require 
explicit shorter-range degrees of freedom, for example $\pi N \to \rho N$. 
Good fits to the $NN$ $^1D_2$ scattering parameters satisfying this 
limitation required that not all $A_n$ be zero. A comparison between the 
phase shift $\delta$ and inelasticity $\eta$ from our best fit and those 
derived from $pp$ scattering analyses \cite{arndt07} is shown in 
Figs.~\ref{fig:del} and \ref{fig:eta}. Here, $\delta$ and $\eta$ are given 
in terms of $S$ and $T$ matrices by $S=1+2iT=\eta\exp(2i\delta)$. A variance 
of 0.02 was used for Re~$T$ and Im~$T$ in these fits. We note that the 
decrease of the inelasticity $\eta$ from a value 1 is due to the $r=2$ $NN'$ 
subchannel which generates the inelastic cut starting at the $\pi NN$ 
threshold, and that no explicit ${\cal D}_{12}$ pole term was introduced 
in the $r=3$ $N\Delta'$ subchannel. 

\noindent 
{\bf Results and discussion.}~~~Applying standard three-body techniques 
\cite{gg11} to the Faddeev equations of our $\pi N \Delta'$ three-body model, 
the following homogeneous integral equation for $T$-matrix poles is obtained: 
\begin{equation}
T_3(W;q_3)={\cal T}_3(W;q_3)\int_0^\infty dq_3^\prime M(W;q_3,q_3^\prime)
T_3(W;q_3^\prime), 
\label{eq:T3} 
\end{equation}  
where, suppressing the dependence on $W$, 
\begin{equation} 
M(q_3,q_3^\prime)=2\int_0^\infty dq_1 K_{31}(q_3,q_1){\cal T}_1(q_1)
K_{13}(q_1,q_3^\prime), 
\label{eq:M3} 
\end{equation} 
\begin{eqnarray} 
K_{31}(W;q_3,q_1) = \frac{1}{2}q_3q_1\int_{-1}^1 d{\rm cos}\theta\,
g_3(p_3)\, g_1^{N\Delta'}(p_1)  &  \nonumber  \\ 
\times \frac{\hat p_3\cdot\hat q_1}
{W-E_1(q_1)-E_2(\vec q_1 + \vec q_3)-E_3(q_3)+i\epsilon}, & 
\label{eq:K} 
\end{eqnarray} 
with $K_{13}(W;q_1,q_3)=K_{31}(W;q_3,q_1)$. The factor 2 on the rhs of 
Eq.~(\ref{eq:M3}) takes into account that the decay ${\cal D}_{12}\to N
\Delta'$ may proceed with either one of the two nucleons in this $\pi NN$ 
dibaryon. The three-vectors $\vec p_3$ and $\vec p_1$ are pair relative 
momenta which in the case of relativistic kinematics are given in terms of 
$q_3$, $q_1$ and cos$\theta$ by Eqs.~(39-43) of Ref.~\cite{gg11}. We note 
that the fictitious baryon $N'$ enters the calculation of the resonance pole 
through the definition of the in-medium propagator of the ${\cal D}_{12}$ 
isobar, ${\cal T}_1(W;q_1)$ of Eq.~(\ref{eq:calT1}). 

The integral equation (\ref{eq:T3}) coincides with that depicted in 
Fig.~\ref{fig:DelDel'} for the $\Delta\Delta'$ eigenvalue problem. In 
practice, replacing $T_3(W;q_3)$ by $Z_3(W;q_3)=T_3(W;q_3)/{\cal T}_3(W;q_3)$, 
Eq.~(\ref{eq:T3}) was transformed to a standard eigenvalue equation 
\begin{equation}
Z_3(W;q_3)=\int_0^\infty dq_3^\prime M(W;q_3,q_3^\prime)
{\cal T}_3(W;q_3^\prime)Z_3(W;q_3^\prime) 
\label{eq:Z3}
\end{equation} 
which was solved numerically. In order to search for resonance poles of 
(\ref{eq:T3}), the integral equation was extended into the complex plane 
using the standard procedure $q_i\to q_i\exp(-i\phi)$ \cite{pearce84} which 
opens large sections of the unphysical Riemann sheet so that one can search 
for eigenvalues of the form $W=M-i\;(\Gamma/2)$. 

\begin{table}[htb] 
\caption{Lowest $\chi^2$/N values of $NN$ $^1D_2$ fits, fitted range 
parameters $\alpha_{1,2,3}$ (in fm$^{-1}$) of the $NN$--$NN'$--$N\Delta'$ 
form factors (\ref{eq:g1}) upon fixing $A_{1,2,3}$ (in fm$^2$), and 
Faddeev-calculated ${\cal D}_{03}$ pole positions $W_k$ (in MeV) 
for $\pi N$ form factors labeled $k$ in Table~\ref{tab:piN}.} 
\begin{ruledtabular} 
\begin{tabular}{ccccccc} 
$A_{1,2,3}$ & $\chi^2$/N & $\alpha_1$ & $\alpha_2$ & $\alpha_3$ & $W_1$ & 
$W_2$  \\ 
\hline 
1,1,1 & 0.78 & 1.47 & 2.27 & 3.24 & $2383-i\;41$ & $2343-i\;24$ \\ 
0,1,1 & 1.10 & 2.00 & 2.11 & 2.96 & $2384-i\;44$ & $2356-i\;30$ \\ 
0,1,$\frac{3}{2}$ & 1.15 & 2.04 & 2.16 & 2.44 & $2392-i\;52$ & 
$2380-i\;45$ \\
\end{tabular}
\end{ruledtabular}
\label{tab:results}
\end{table}

In the actual solution of the eigenvalue equation (\ref{eq:Z3}) we replaced 
the Breit-Wigner mass value $m_{\Delta'}$=1232~MeV in the propagator 
${\cal T}_3$, Eq.~(\ref{eq:calT3}), by an effective $\Delta$-pole complex 
mass value \cite{arndt06} $m_{\Delta}=1211-i\;(2/3)\times 49.5$~MeV, 
where the 2/3 suppression factor accounts for the $\Delta$ decay phase 
space effective in the $I(J^P)=0(3^+)$ $\pi N\Delta$ three-body system, 
as discussed earlier. ${\cal D}_{03}$ resonance pole positions calculated 
using the three lowest $\chi^2$ fits of the $NN$--$NN'$--$N\Delta'$ form 
factors (\ref{eq:g1}) to the $^1D_2$ $NN$ scattering parameters are listed 
in Table~\ref{tab:results}, for the two $\pi N$ form factors specified in 
Table~\ref{tab:piN}. 
Comparing the calculated mass values Re$\;W$ in the last column to those 
in the preceding column, one observes sensitivity to the spatial extension 
of the $\pi N$ form factor, quantified by the values of $r_0$ from 
Table~\ref{tab:piN} which we consider as providing reasonable bounds 
on this spatial extension. The smaller $r_0$, the lower the calculated 
mass values are. Admitting values of $r_0$ appreciably below 0.9~fm requires 
the introduction of explicit vector-meson and/or quark-gluon degrees of 
freedom. Within a given column for $W$ in the table, the calculated mass 
values display sensitivity to the $N\Delta'$ form factor primarily through 
the fitted values of $\alpha_3$ listed here. The larger $\alpha_3$, the lower 
the calculated mass values are. In general, it was found impossible to get 
values of $\alpha_3 \lesssim 2.5$~fm$^{-1}$, whereas going beyond $\alpha_3 
\sim 3$~fm$^{-1}$ was considered undesirable, again requiring the introduction 
of explicit short-range degrees of freedom. As for width values $-2\;{\rm Im}
\;W$, the calculated widths display little sensitivity to these form factors 
and the widths are determined mostly by the phase space available for decay. 
Averaging over the pole positions of the best-fit solutions in the first row 
of Table~\ref{tab:results}, the mass and width values for ${\cal D}_{03}$ are 
\begin{equation} 
{\cal D}_{03}: \;\;\; M=2363\pm 20,\;\;\; \Gamma=65\pm 17 \;\;\; ({\rm MeV}). 
\label{eq:mass-width} 
\end{equation} 
By scaling to the mass-width phenomenology of the known $\Delta$ resonance 
\cite{arndt06}, the corresponding ${\cal D}_{03}$ Breit-Wigner mass-width 
values for comparison with experiment \cite{bash09} should each be about 10 
MeV higher.   

In summary, considering the $\Delta$ resonance as a stable baryon $\Delta'$, 
we have presented a dynamical $\pi N \Delta'$ separable-potential model for 
the ${\cal D}_{03}$ dibaryon that captures the essential physics of the 
underlying pairwise interactions, using fitted form factors for the $p$-wave 
$\pi N$ and the $s$-wave $N\Delta'$ interactions in the channels dominated 
by the $\Delta$(1232) baryon resonance and the ${\cal D}_{12}(2150)$ dibaryon 
resonance, respectively. The corresponding three-body Faddeev equations 
were derived and the resulting effective two-body $\Delta\Delta'$ 
eigenvalue equation of Fig.~\ref{fig:DelDel'} was solved, replacing the 
$\Delta'$-spectator real mass in the in-medium $\pi N$ propagator by 
a physical $\Delta$ effective complex mass value. A robust ${\cal D}_{03}$ 
dibaryon pole was found above the $\pi{\cal D}_{12}$ threshold but below the 
$\Delta\Delta$ threshold, with mass value $M=(2.36\pm 0.02)$~GeV in good 
agreement with the location of the $pn\to d\pi\pi$ resonance observed by 
WASA@COSY. Furthermore, the calculated width of $\Gamma=65\pm 17$~MeV also 
agrees well with the observed width of 70 MeV. 

\noindent 
{\bf Acknowledgments}:~~H.G. is supported in part by COFAA-IPN (M\'exico) 
and A.G. by the HadronPhysics3 networks SPHERE and LEANNIS of the European 
FP7 initiative.

\end{document}